\documentclass[floatfix,pra,twocolumn,showpacs,amsmath,amssymb,letterpaper,groupaddresses,superscriptaddress]{revtex4}
\usepackage{times}
\usepackage{latexsym}
\usepackage{graphicx}
\usepackage{verbatim,times,bbm}
\usepackage{color}

\newcommand{\dint}{{\int \!\!\!\! \int}}

\usepackage{color}

\begin{document}

\title{Capacities of linear quantum optical systems}

\author{Cosmo Lupo}
\affiliation{School of Science and Technology, University of Camerino, I-62032 Camerino, Italy}

\author{Vittorio Giovannetti}
\affiliation{NEST, Scuola Normale Superiore and Istituto Nanoscienze-CNR, I-56126 Pisa, Italy}

\author{Stefano Pirandola}
\affiliation{Department of Computer Science, University of York, York YO10 5GH, UK}

\author{Stefano Mancini}
\affiliation{School of Science and Technology, University of Camerino, I-62032 Camerino, Italy}
\affiliation{INFN-Sezione di Perugia, I-06123 Perugia, Italy}

\author{Seth Lloyd}
\affiliation{Department of Mechanical Engineering, MIT Cambridge, MA 02139, USA}

\begin{abstract}

A wide variety of communication channels employ the quantized 
electromagnetic field to convey information. 
Their communication capacity crucially depends on losses associated 
to spatial characteristics of the channel such as diffraction and 
antenna design. 
Here we focus on the communication via a finite pupil, showing that 
diffraction is formally described as a {\it memory channel}.
By exploiting this equivalence we then compute the communication 
capacity of an optical refocusing system, modeled as a converging lens.
Even though loss of information originates from the finite pupil 
of the lens, we show that the presence of the refocusing system can 
substantially enhance the communication capacity.
We mainly concentrate on communication of classical information,
the extension to quantum information being straightforward.

\end{abstract}


\maketitle

\section{Introduction}

The most prominent candidate for implementing long
distance quantum communication is undoubtedly represented by the
electromagnetic field (EMF)~\cite{longd}. Although quantum information
theory is more commonly represented in terms of discrete variables
(e.g., qubits), information is most naturally encoded in the EMF
 by means of continuous variables, which, in the quantum
domain, are described by bosonic degrees of freedom. Moreover, all
the fundamental quantum information tools and protocols have been
demonstrated for continuous variable systems~\cite{review1,review2}, 
from quantum computation~\cite{tools} to quantum error
correction~\cite{tools0,tools1}, quantum teleportation~\cite{tools2} 
and quantum key distribution~\cite{tools4,tools5,tools6}. 
Here we consider the problem of optical quantum
communication~\cite{QOchannel,Shapiro,YUENSH}, and compute the
communication capacity, i.e., the maximum rate at which
information can be reliably transmitted. 
Although we explicitly consider communication of classical 
information~\cite{Ccapacity}, our results are immediately 
extensible to the case of quantum information~\cite{Qcapacity}.

The most general and simple, although physically relevant,
mathematical model of optical communication line is the Gaussian
channel~\cite{HolWer}, which describes the linear propagation of the EMF. 
In the classical domain, the ultimate limits for
communication via Gaussian channels were provided by the seminal
work of Claude E.~Shannon~\cite{Shannon}. In the quantum domain,
the structure of Gaussian channels is notably
rich~\cite{Gchannels}, with non-trivial properties in terms of
degradability~\cite{GchannelsDEG} and
security~\cite{GchannelsSEC,GchannelsSEC2}. However, a full
information theoretical characterization has been presently
achieved only for certain families of channels, such as the lossy
channel~\cite{broadband,Wolf,PS}. These results have been applied to
compute the maximum rate of reliable communication via attenuating
media, as optical fibers, wave-guides, and via free-space
propagation~\cite{freespace,Shapiro,wave}. 
Moving along this line we shall provide the information 
theoretical description of the effects of the signal propagation 
through lossy communication channels with linear characteristic.
After introducing the general methods, we consider the example
of an optical refocusing system with finite pupil, which is 
schematized as a thin lens which is placed between the sender of 
the message and the receiver under focusing conditions.
Notwithstanding its relatively simple structure this setup
captures the basics features of all those situations in which a
transmitted signal is either focused on a detector by a suitable
optical system prior to the information decoding process, or where
it  has to be refocused by a suitable repeater to allow long
distance communication, e.g., by means of parabolic antenna for
satellite communication~\cite{satellite}.

We explictly discuss how the signal diffraction through the optical
system can be formally described as a quantum channel with correlated 
noise ({\it memory channel})~\cite{correlated}, where correlations are 
quantified in terms of the associated Rayleigh length.
Therefore the information theoretical characterization of the resulting 
quantum channel is carried out using tools and methods that have been 
previously applied to quantum memory channels~\cite{unravel}.
In this framework the main effect of the signal propagation through the optical
system is to introduce the diffraction of light caused by its finite
pupil, leading to bandwidth limited communication~\cite{telegraph}.
However, when compared with the free-space communication scheme~\cite{freespace}, 
the presence of the refocusing apparatus may yield an improvement in the channel 
capacity of the system. 
This possibility has been put forward in~\cite{OptRef}.
Here we provide a detailed derivation of the channel model together with 
the analysis of results in several configurations. 
Furthermore, we also extend the approach to encompass the case of non-monochromatic light.

The article proceeds as follows. 
In Sec.~\ref{Linear}, we provide the general model for communication 
via lossy multimode quantum optical channels. 
In Sec.~\ref{TheSystem}, we introduce a simplest example of such a 
system, communication through a converging lens with finite pupil, 
and show that it can be formally described as a channel with correlated noise.
In Sec.~\ref{TheChannel} we quantize this system, whose communication 
capacity is computed in Sec.~\ref{Mono} for the case of monochromatic light. 
We then compare the communication via the refocusing system with
the free-space setting in Sec.~\ref{Refocus}.
Section~\ref{NonMono} is devoted to the extension to the 
non-monochromatic case, and Sec.~\ref{TheEnd} is for final remarks.

\section{Communication capacity of quantized linear optical systems}\label{Linear}

Consider a linear optical system with a set of $M$ transmitter modes,
labelled by $i$, and $N$ receiver modes, labelled by $j$.  
In the case of radio-frequency or microwave communication, for example,
the transmitter and receiver could be antennae. 
For optical communication, the transmitter could be a laser coupled to 
a telescope, and the receiver could be a telescope coupled
to a charge-coupled device (CCD) array. 
Transmitter and receiver modes typically have both spatial 
characteristics determined by the optical properties of the setup, and temporal characteristics, 
determined by the frequency and bandwidth of the radiation employed in
the communication.  
The transmittivity matrix $T_{ji}$ gives the
fraction of light from the $i$-th transmitter mode that is
received at the $j$-th receiver mode.  We would like to
determine the maximum amount of information that can
be sent for fixed total
input power.

First, consider the purely lossy case, in which noise from
the environment is neglible. This is the case, for example,
for free-space optical communication in a thermal background.
The addition of noise will be considered below.
When there is just one transmitter mode and one receiver
mode, the channel is simply the lossy bosonic channel,
whose classical capacity is known~\cite{broadband}: if the loss is $\eta$
and the single-use average photon number is $\bar n$,
then the number of {\it bits} that can be sent down the channel
is $g(\eta\bar n)$, where 
\begin{eqnarray}\label{gfunction}
g(x) := \left\{ \begin{array}{lr}
(x+1) \log_2 (x+1) - x \log_2 x & \mbox{for} \ \ x > 0 \, , \\
0                               & \mbox{for} \ \ x \leqslant 0 \, .
\end{array}\right.
\end{eqnarray}
The capacity is attained by sending coherent states down the channel. 
If there are $M$ parallel channels, with
average photon number $\bar n_i$ for the $i$-th channel,
then the capacity is simply the sum of the capacities for
the individual channels.

In our case, we have a single multimode lossy channel with
transmittivity matrix $T_{ji}$, which mixes the
input modes together. This channel can be transformed
into a set of parallel channels by using the singular value decomposition. 
The singular value decomposition states that
any $N\times M$ matrix $T$ can be written as
\begin{equation}\label{defEqT}
T = {\cal V} \Sigma {\cal U} \, ,
\end{equation}
where ${\cal V}$ is an $N\times N$ unitary matrix, 
$\Sigma$ is an $N\times M$ matrix with entries only
on the diagonal, and ${\cal U}$ is an $M\times M$ unitary matrix. 
In our case, we can write the transmittivity matrix $T$ in
components as
\begin{equation}
T_{ji} = \sum_k {\cal V}_{jk} \sqrt{\eta_k} {\cal U}_{ki} \, ,
\end{equation}
where $\{ \sqrt{\eta_k} \}$ are the singular values of the transmittivity
matrix. The singular value decomposition shows that any multimode
lossy channel can be decomposed into parallel lossy channels
with input modes corresponding to the rows of ${\cal U}$, 
output modes corresponding
to the columns of ${\cal V}$, and loss factors corresponding to
the singular values $\eta_k$.

The singular value input modes 
can now be quantized using annihilation and creation
operators $a_j, a^\dagger_j$: $[a_j, a^\dagger_{j'}] = \delta_{j,j'}$.
Similarly, the output modes can be quantized using annihilation and creation
operators $b_j, b^\dagger_j$: $[b_j, b^\dagger_{j'}] = \delta_{j,j'}$.
To preserve the canonical quantization relationships, each
input-output pair is coupled to a environment mode with annihilation
and creation operators $\xi_j, \xi^\dagger_j$:
\begin{equation} 
b_j = \sqrt{\eta_j} \, a_j + \sqrt{1-\eta_j} \, \xi_j \, .
\end{equation}
We see that the singular value decomposition of the multimode
lossy quantum channel renders the channel completely equivalent
to a set of parallel lossy quantum channels with loss factors
$\eta_j$.  The communication capacity of the channel when
$\bar n_j$ photons are transmitted down the $i$-th singular
value mode is simply $\sum_j g(\eta_j \bar n_j)$.

If the $j$-th singular value mode has average energy per
photon $\hbar \bar\omega_j$, then the capacity of the channel
with total energy $E$ per use
is obtained by solving the constrained maximization problem
with Lagrangian
\begin{equation}
\sum_j g\left( \eta_j \bar n_j \right) - \mu \left( \sum_j \bar n_j \hbar \bar \omega_j - E \right) \, ,
\end{equation} 
yielding the solution
\begin{equation}
\bar{n}_j = \left[ \eta_j \left( 2^{\mu \hbar \bar{\omega}_j/\eta_j} - 1 \right) \right]^{-1}
\end{equation}
Here, the Lagrange multiplier $\mu$ is chosen to give
the proper total energy $E$.
In the following, when dealing with monochromatic light, we 
express the energy in terms of number of photons, $\bar n = E/(\hbar \omega)$.

Thus, the singular value decomposition allows us to transform any linear,
lossy, multimode channel with transmittivity matrix $T_{ji}$
into a set of parallel lossy channels.
Quantization then yields the capacity of this set of parallel
channels. We now apply this result to various simple
optical settings. 
In particular, we show that the resulting capacity can
be significantly larger than the capacity which is achieved
by using `naive' coding and decoding techniques.

\section{The optical system}\label{TheSystem}

\begin{figure}
\centering
\includegraphics[width=0.4\textwidth]{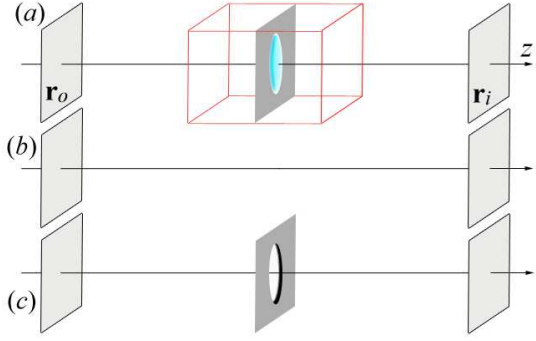}
\caption{(Color online.) ($a$) Optical communication
through an optical refocusing apparatus, modeled as a thin lens
of radius $R$ and focal lenght $f$.
($b$) The free-space propagation scenario.
($c$) An alternative scenario in which the lens is replaced by a
hole of the same size in the absorbing screen.
$\mathbf{r}_o$ and $\mathbf{r}_i$ denote the Cartesian
coordinates on the object and image planes, respectively.} \label{optical}
\end{figure}

We consider the propagation of light along an optical axis~$z$. 
The input and output signals are identified by the transverse light 
fields at two planes orthogonal to the optical axis, respectively 
identified as the object plane and image plane, see Fig.~\ref{optical}(a). 
In classical scalar optics, the monochromatic optical fields at the object and
image plane are described by scalar functions $U_o(\mathbf{r}_o)$,
$U_i(\mathbf{r}_i)$, where $\mathbf{r}_o$ and $\mathbf{r}_i$ are the
cartesian coordinates at the object and image plane. For linear
systems, including the free-space propagation of light, the input/output
relations at the object and image planes, respectively, are
described by a transfer function $T(\mathbf{r}_i,\mathbf{r}_o)$,
such that
\begin{equation}\label{IO}
U_i(\mathbf{r}_i) = \dint d^2\mathbf{r}_o \,
T(\mathbf{r}_i,\mathbf{r}_o) U_o(\mathbf{r}_o) \, .
\end{equation}
The quantum version of the relations (\ref{IO}) can be derived by
applying the canonical quantization procedure and introducing a
proper set of normal modes for the input and output fields (as it is
done in~\cite{Shapiro} for the free-space propagation).

As mentioned in the introduction we model the optical refocusing system as a
converging lens of focal length $f$, located at distance $D_o$ from
the object plane. Working in the thin lens, paraxial approximation, 
and neglecting aberrations, light is collected at the image plane located 
at distance $D_i$ from the optical system, where $1/D_o + 1/D_i = 1/f$. 
Eventually the image is magnified by a factor $M = D_i/D_o$. 
Diffraction of light is responsible for image blurring and causes loss of 
information. 
It can be described by introducing an effective entrance pupil
characterizing the optical system. Denoting $P(\mathbf{r})$ the
characteristic function of the pupil that encircles the lens, the 
transfer function for the monochromatic field of wavelength $\lambda$ 
is obtained by Fourier transforming~\cite{Goodman}:
\begin{eqnarray}\label{transfer}
T(\mathbf{r}_i,\mathbf{r}_o) =
\tfrac{e^{\iota\vartheta(\mathbf{r}_i,\mathbf{r}_o)}}{\lambda^2 D_o D_i} \dint d^2 \mathbf{r} \, P(\mathbf{r})\;  e^{-\iota 2\pi
\frac{(\mathbf{r}_i-M\mathbf{r}_o)}{\lambda D_i} \cdot\mathbf{r}} \, ,
\end{eqnarray}
(throughout the paper $\iota$ denotes the imaginary unit).
In writing the above expression we are implicitly assuming that the light rays which do not 
hit the pupil will not reach the image plane (either because they are scattered 
out, or because they are absorbed by some medium). 
Such hypothesis is not fundamental and could be dropped by adding an extra 
term to Eq.~(\ref{transfer}) which accounts for the diffraction of the 
light rays that miss the pupil.
Eventually, we notice that, in the paraxial approximation, 
the acquired phase results
\begin{equation}
\vartheta(\mathbf{r}_i,\mathbf{r}_o) = \frac{\pi}{\lambda D_o}
\left( |\mathbf{r}_o|^2 + |\mathbf{r}_i|^2/M \right) + \frac{2\pi
D_o}{\lambda}\left( 1+M \right) \, .
\end{equation}
For instance, in the case of a circular lens of radius $R$, the
pupil function is
\begin{eqnarray}\label{pupil}
P(\mathbf{r}) = \left\{\begin{array}{ccc}
1 & \mbox{for} & |\mathbf{r}| < R \, ,\\
0 & \mbox{for} & |\mathbf{r}| > R \, ,
\end{array}\right.
\end{eqnarray}
and the transfer function reads
\begin{equation}
T(\mathbf{r}_i,\mathbf{r}_o) =
\frac{e^{\iota\vartheta(\mathbf{r}_i,\mathbf{r}_o)} R^2}{\lambda^2 D_o
D_i} \frac{J_1(2\pi R \rho )}{R\rho} \, ,
\end{equation}
where $J_1$ indicates the Bessel function of first kind and order
one, and $\rho := |\mathbf{r}_i-M\mathbf{r}_o|/(\lambda D_i)$.

\subsection{Light diffraction as a memory channel}\label{Memory}

We shall see that the effects of diffraction on light propagation
can be described as memory effects in the communication channel. Let
us recall that a {\it memoryless} channel is such that its action on
different channel inputs are identical and independent. Conversely,
the actions of a {\it memory channel}, also called a {\it channel
with correlated noise}, at different uses are not independent and/or
not identical.

Let us assume that information is encoded at the object plane by an
array of {\it pixels} located at positions $\mathbf{r}_o(k)$, with
the integer $k$ labeling the pixels. Different signals are emitted
from different pixel positions, which play the role of an array of
independent channel inputs. It follows from Eq.~(\ref{IO}) that the
input from the $k$-th pixel is mapped to an output field at the
image plane with spatial amplitudes
\begin{equation}
U_i^{(k)}(\mathbf{r}_i) = T(\mathbf{r}_i,\mathbf{r}_o(k)) \,
U_o^{(k)} \, .
\end{equation}
Let us notice that, even though the action of the channel is
identical for all the input pixels, the output fields are not
mutually independent. In fact, a pair of output fields,
corresponding to the the $k$-th and $k'$-th inputs have a non
vanishing spatial overlap
\begin{equation}
\mathcal{C}_{k,k'} := \frac{\dint d^2\mathbf{r}_i \,
U_i^{(k)}(\mathbf{r}_i) U_i^{(k')*}(\mathbf{r}_i)}{U_o^{(k)}
U_o^{(k')*} } \, ,
\end{equation}
that is,
\begin{equation}
\mathcal{C}_{k,k'} = \dint d^2\mathbf{r}_i \,
T(\mathbf{r}_i,\mathbf{r}_o(k)) T^*(\mathbf{r}_i,\mathbf{r}_o(k')) \, .
\end{equation}
The overlap between output signals may cause interference, which in
turn produces distortion and loss of information in the
communication via the optical channel. In particular, the overlap
with the output field generated by the $k$-th input signal induces
noise in the detection of the $k'$-th output. If the overlap between
the two output fields is not negligible --- i.e., diffraction
produces sensible effects --- the noise affecting the $k'$-th 
output field turns out to be highly correlated with the input signal at the $k$-th
pixel.

Introducing the dimensionless variable $\tilde{\mathbf{r}} :=
\mathbf{r}/R$ (here $R$ is the linear extension of the entrance
pupil) and using the expression in Eq.~(\ref{transfer}) for the
transfer function, we get the following expression for the output
signal overlap
\begin{eqnarray}
\mathcal{C}_{k,k'} = \tfrac{e^{\iota\delta\vartheta}}{x_\mathsf{R}^2}
\dint d^2\tilde{\mathbf{r}} \, P(R\tilde{\mathbf{r}}) e^{-\iota
2\pi\frac{(\mathbf{r}_o(k)-\mathbf{r}_o(k'))}{x_\mathsf{R}} \cdot
\tilde{\mathbf{r}} } \, ,
\end{eqnarray}
from which it is evident that the spatial overlap between the two
output fields is determined by the Rayleigh length of the apparatus
\begin{eqnarray}\label{rayleigh}
x_\mathsf{R} = \frac{\lambda D_o}{R} \;.
\end{eqnarray}
For instance, in the case of a circular pupil
of radius $R$, we obtain
\begin{eqnarray}
\mathcal{C}_{k,k'} = \frac{e^{\iota\delta\vartheta}}{x_\mathsf{R}}
\frac{J_1(2\pi|\mathbf{r}_o(k)-\mathbf{r}_o(k')|/x_\mathsf{R})}{|\mathbf{r}_o(k)-\mathbf{r}_o(k')|} \, .
\end{eqnarray}
In conclusion the Rayleigh length, which quantifies the amount of
diffraction in the optical system, also quantifies the degree of
correlations in the optical communication channel. The
corresponding quantum memory channel manifests {\it inter-symbol
interference} effects, and is qualitatively analogous to those
studied in~\cite{unravel,SMPRA}.

\section{The quantum channel}\label{TheChannel}

The light field at the object and image planes can be quantized
according to standard canonical quantization. In order to derive the
quantum version of the input/output relations in Eq.~(\ref{IO}) we
first identify a proper set of normal modes at the input and output field 
and proceed along the lines detailed in Sec.~\ref{Linear}.

We assume that information is encoded in the object plane on a square of length
$L$, creating an image on the image plane which (in the geometric
optics approximation) is contained in a square of size $ML$. We
hence introduce the field variables
\begin{eqnarray}
U_o(\mathbf{n}_o) & := & \frac{1}{L} \dint d^2\mathbf{r}_o \,
e^{-\iota 2\pi \left(\frac{\mathbf{n}_o\cdot\mathbf{r}_o}{L} - \frac{|\mathbf{r}_o|^2}{2\lambda D_o}-\frac{D_o}{\lambda}\right)} 
U_o(\mathbf{r}_o) \, ,\nonumber\\
U_i(\mathbf{n}_i) & := & \frac{1}{ML} \dint d^2\mathbf{r}_i \, 
e^{-\iota 2\pi \left( \frac{\mathbf{n}_i\cdot\mathbf{r}_i}{ML} +\frac{|\mathbf{r}_i|^2}{2\lambda D_i} + \frac{D_i}{\lambda}\right)}
U_i(\mathbf{r}_i) \, , \nonumber \\
\label{Fourier_ab}
\end{eqnarray}
where the integral over $\mathbf{r}_o$ is restricted to the surface 
of area $L\times L$ which encircles the object, while the integral 
over $\mathbf{r}_i$ is restricted to the surface of area $ML \times ML$ 
which defines the receiving screen, and where $\mathbf{n}_o$ and 
$\mathbf{n}_i$ are vectors having two integer components.
The functions $U_o(\mathbf{n}_o) $ and $U_i(\mathbf{n}_i)$ express 
the field components of transverse momentum $2\pi \mathbf{n}_o/L$ and 
$2\pi \mathbf{n}_i/(ML)$, respectively at the object and image plane. 
Substituting Eqs.~(\ref{Fourier_ab}) into (\ref{IO}), and using 
(\ref{transfer}), we write the input/output relations in the form
\begin{equation}
U_i(\mathbf{n}_i) = \sum_{\mathbf{n}_o} \,
T_{\mathbf{n}_i,\mathbf{n}_o} \, U_o(\mathbf{n}_o) \, ,
\end{equation}
where the transfer matrix $T_{\mathbf{n}_i,\mathbf{n}_o}$ reads
\begin{equation}\label{Tmatrix}
T_{\mathbf{n}_i,\mathbf{n}_o} = \frac{1}{\lambda^2 D_o D_i} \dint
d^2\mathbf{r} \, P(\mathbf{r})
\Delta_{\mathbf{n}_i,\mathbf{n}_o}(\mathbf{r}) \, ,
\end{equation}
with
\begin{align}
\Delta_{\mathbf{n}_i,\mathbf{n}_o}(\mathbf{r}) = & \frac{1}{ML^2} 
\dint d^2\mathbf{r}_o \, e^{\iota 2\pi \left(\frac{\mathbf{n}_o}{L}+\frac{M\mathbf{r}}{\lambda D_i}\right) \cdot \mathbf{r}_o} \nonumber \\
& \times \dint d^2\mathbf{r}_i \, e^{ - \iota 2\pi \left(\frac{\mathbf{n}_i}{ML}+\frac{\mathbf{r}}{\lambda D_i}\right)
\cdot \mathbf{r}_i } \, . \label{Deltamatrix}
\end{align}
Finally, 
we consider the {\it singular value decomposition} of the
transfer matrix,
\begin{equation}\label{SVD}
T_{\mathbf{n}_i,\mathbf{n}_o} = \sum_{\mathbf{n}} \,
\mathcal{V}_{\mathbf{n}_i,\mathbf{n}} \, \sqrt{\eta_{\mathbf{n}}} \,
\mathcal{U}_{\mathbf{n},\mathbf{n}_o} \, ,
\end{equation}
where, as in Eq.~(\ref{defEqT}),
$\mathcal{U}_{\mathbf{n},\mathbf{n}_o}$,
$\mathcal{V}_{\mathbf{n}_i,\mathbf{n}}$ are unitary matrices, and
$\{ \sqrt{\eta_{\mathbf{n}}} \}$ are the {\it singular values} 
of the matrix $T_{\mathbf{n}_i,\mathbf{n}_o}$, taking values in the interval $[0,1]$. 
A set of input and output field variables are hence defined as follows:
\begin{eqnarray}
\tilde U_o(\mathbf{n}) & := & \sum_{\mathbf{n}_o} \, \mathcal{U}_{\mathbf{n},\mathbf{n}_o} \, U(\mathbf{n}_o) \, , \\
\tilde U_i(\mathbf{n}) & := & \sum_{\mathbf{n}_i} \,
\mathcal{V}^*_{\mathbf{n}_i,\mathbf{n}} \, U_i(\mathbf{n}_i) \, .
\end{eqnarray}
It follows from Eq.~(\ref{SVD}) that they satisfy the identities
\begin{equation}\label{classical}
\tilde U_i(\mathbf{n}) = \sqrt{\eta_{\mathbf{n}}} \, \tilde
U_o(\mathbf{n}) \, .
\end{equation}
In other words, the field variables $\tilde U_o(\mathbf{n})$ are
independently, but non necessarily identically, transmitted to the
output variables $\tilde U_i(\mathbf{n})$. The effect of the channel
is to attenuate the $\mathbf{n}$-th variable by a factor
$\eta_\mathbf{n}$.

We can now promote the output and input field variables to the rank
of quantum operators, by substituting
\begin{align}
\tilde{U}_o(\mathbf{n}) & \rightarrow \sqrt{\hbar\omega/2} \, 
a_\mathbf{n} \, , \quad \tilde{U}^*_o(\mathbf{n}) \rightarrow 
\sqrt{\hbar\omega/2} \, a^\dag_\mathbf{n} \, , \\
\tilde{U}_i(\mathbf{n}) & \rightarrow \sqrt{\hbar\omega/2} \,
b_\mathbf{n} \, , \quad \tilde{U}^*_i(\mathbf{n}) \rightarrow
\sqrt{\hbar\omega/2} \, b^\dag _\mathbf{n} \, ,
\end{align}
where $\omega = 2\pi c/ \lambda$ is the frequency, and imposing the
canonical commutation relations:
\begin{align}
[ a_{\mathbf{n}} , a_{\mathbf{n}'}^\dag ] & =
\delta_{\mathbf{n},\mathbf{n}'} \, , \\
{[} b_{\mathbf{n}} , b_{\mathbf{n}'}^\dag {]} & =
\delta_{\mathbf{n},\mathbf{n}'} \, .
\end{align}
The preservation of canonical commutation relations requires to
invoke a set of canonical noise variables $\{ \xi_\mathbf{n} ,
\xi_\mathbf{n}^\dag \}$ and to write the quantum version of
Eq.~(\ref{classical}) as
\begin{equation}\label{quantum}
b_\mathbf{n} = \sqrt{\eta_\mathbf{n}} \, a_\mathbf{n} +
\sqrt{1-\eta_\mathbf{n}} \, \xi_\mathbf{n} \, .
\end{equation}
This set of input/output relations, together with their hermitian
conjugates, characterizes, upon evaluation of the transmissivities
$\eta_\mathbf{n}$, the quantum description of the optical channel.

\section{Capacity of the optical quantum communication channel}\label{Mono}

The input/output relations for the quantum description of the
channel which have been derived in Eq.~(\ref{quantum}) formally
qualify the optical system as a {\it broadband lossy channel}: a
multimode channel in which a collection of bosonic modes is
transmitted with corresponding efficiencies. This model has been
characterized from the information-theoretical viewpoint in~\cite{broadband,Wolf}, 
where the capacities of the channel for transmitting classical and 
quantum information have been established.

The number of modes which are transmitted through the channel is
virtually infinite. Actually, due to the finiteness of the entrance
pupil, only a finite number of modes has nonzero transmissivity. 
The values of the transmissivities $\eta_\mathbf{n}$ can be numerically
estimated by first computing the elements of the transfer matrix
$T_{\mathbf{n}_i,\mathbf{n}_o}$, and then its singular values.
Indeed, the matrix in Eq.~(\ref{Deltamatrix}) can be rewritten in
terms of the ratio $L/x_\mathsf{R}$, and an analytical solution for
the values of the effective transmissivities can be deduced in the
following limits, namely
\begin{eqnarray}
&&L \ll x_\mathsf{R} \;,   \quad \qquad \quad  \mbox{FAR FIELD,} \label{FARFIE} \\ \nonumber \\
&&L \gg x_\mathsf{R} \;,   \quad \qquad \quad  \mbox{NEAR FIELD.}  \label{NEARFIE}
\end{eqnarray} 

First of all, we notice that in the far-field case
Eq.~(\ref{Deltamatrix}) reads:
\begin{equation}
\Delta_{\mathbf{n}_i,\mathbf{n}_o} \simeq M L^2 \,
\delta_{\mathbf{n}_i,\mathbf{0}} \, \delta_{\mathbf{n}_o,\mathbf{0}} \, ,
\end{equation}
where $\delta$ denotes the Kronecker symbol. 
This expression in turn leads to
\begin{equation}
T_{\mathbf{n}_i,\mathbf{n}_o} \simeq \pi \left(
\frac{L}{x_\mathsf{R}} \right)^2 \, \delta_{\mathbf{n}_i,\mathbf{0}}
\, \delta_{\mathbf{n}_o,\mathbf{0}} \, ,
\end{equation}
that is, in the far-field regime only one mode is transmitted
through the optical channel, and it is attenuated by a factor
$\simeq \pi^2 ( L/x_\mathsf{R} )^4$.  At least for optical frequencies it is reasonable to assume that
the noise modes in Eq.~(\ref{quantum}) are not populated. Under this assumption, using the result of \cite{broadband}, we are led
to the following expression for the classical capacity of the
optical channel:
\begin{equation}\label{farfield}
C_\mathrm{ff} = g\left( \frac{\pi^2 L^4}{x_\mathsf{R}^4} \, \bar n \right) \, ,
\end{equation}
where $\bar n$ denotes the mean number photons at the object plane. 
The above calculations refer to the case of scalar field.
Polarization can be included by doubling  the total number of modes.
In this case Eq.~(\ref{farfield}) will then be replaced by
\begin{equation}
C_\mathrm{ff}^{pol.} = 2 \, g\left( \frac{\pi^2 L^4}{2 x_\mathsf{R}^4} \, \bar n\right) \, , 
\end{equation}
where the factor $2$ which multiplies the $g$-function comes from the
fact that now there are two modes which can efficiently propagate,
while the extra factor $1/2$ in the argument of $g$ comes from the
fact that the available energy must be equipartitioned between them.
Analogously, the quantum capacity can be computed according to~\cite{Wolf}.

Let us now consider the near-field limit, $L \gg x_\mathsf{R}$. In
this regime we can approximate
\begin{equation}
\Delta_{\mathbf{n}_i,\mathbf{n}_o}(\mathbf{r}) \simeq \lambda^2 D_o
D_i \, \delta_{\mathbf{n}_i,\mathbf{n}_o} \,
\delta^{(2)}\left(\mathbf{r}+\frac{\lambda D_o
\mathbf{n}_o}{L}\right) \, ,
\end{equation}
where $\delta^{(2)}$ indicates the two-dimensional Dirac function.
It follows that
\begin{equation}
T_{\mathbf{n}_i,\mathbf{n}_o} \simeq
\delta_{\mathbf{n}_i,\mathbf{n}_o} \, P\left( \frac{\lambda D_o}{L}
\, \mathbf{n}_o\right) \, .
\end{equation}
Within this approximation the transfer matrix is diagonal, hence the
singular values coincide with its diagonal entries. For a circular
pupil of radius $R$, we obtain the following values for the
transmittivities:
\begin{eqnarray}
\eta_{\mathbf{n}} \simeq \left\{\begin{array}{ccc}
1 & \mbox{for} & |\mathbf{n}| < L/x_\mathsf{R} \, , \\
0 & \mbox{for} & |\mathbf{n}| > L/x_\mathsf{R} \, .
\end{array}\right.
\end{eqnarray}
We deduce that, in the near-field regime, the number of transmitted
modes per surface unit is approximatively equal to
$\pi/x_\mathsf{R}^2$, each being transmitted with approximatively
unit efficiency. Equivalently we can say that  the total number of
modes which are transferred with unit efficiency is equal to
\begin{equation}\label{PU}
\nu \simeq \pi (L/x_\mathsf{R})^2 \;. 
\end{equation}
We are now in the condition of computing the capacity.
Denoting by $\bar n$ the average number of photons impinging on the surface,
the capacity has the form \cite{notafs}
\begin{equation}\label{nearfield}
C_\mathrm{nf} = \nu \; g( \bar{n}/\nu) = \frac{\pi L^2}{x_\mathsf{R}^2} \,
g \left( \frac{x_\mathsf{R}^2}{\pi L^2} \, \bar{n} \right) \, ,
\end{equation}
where we used the fact that the maximum transfer is obtained when
$\bar{n}$ is equipartitioned among all transmitted modes. 
It is worth stressing that, since the modes employed in the transmission 
are perfectly transmitted, the classical capacity (expressed in {\it bits})
coicides with the quantum capacity (expressed in {\it qubits}). 
As before, our result can be generalized to
include also the polarization degree of freedom obtaining
\begin{equation}
C_\mathrm{nf}^{pol.} = \frac{2 \pi L^2}{x_\mathsf{R}^2} \, 
g \left( \frac{x_\mathsf{R}^2}{2 \pi L^2} \, \bar{n} \right) \, .
\end{equation}

\begin{figure}[hbt]
\centering
\includegraphics[width=0.4\textwidth]{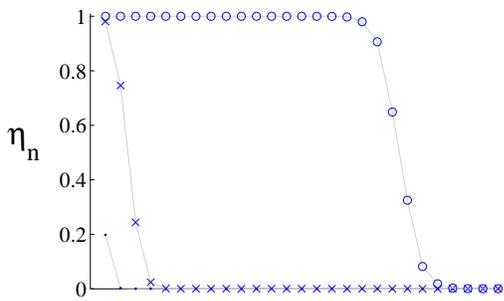}
\caption{The effective transmissivities $\eta_\mathbf{n}$, for
different value of the ratio $L/x_\mathsf{R}$. Dots: $L/x_\mathsf{R}
= 0.1$ (far-field). Crosses: $L/x_\mathsf{R} = 1$.
Circles: $L/x_\mathsf{R} = 10$ (near-field). The figure refers
to a one-dimensional setting, in which information is encoded at the
object plane along an infinite strip of size $L$, and diffraction is
caused by an infinitely long slit of size $2R$.}
\label{transmissivities}
\end{figure}

\begin{figure}[hbt]
\centering
\includegraphics[width=0.4\textwidth]{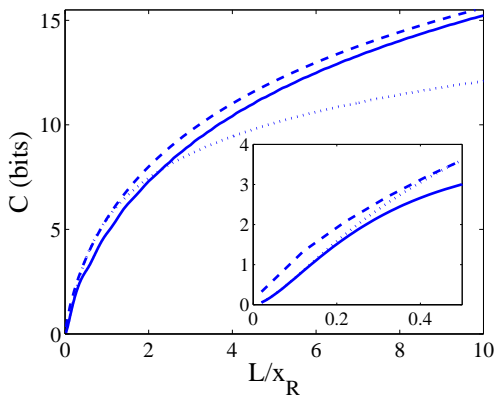}
\caption{The plot shows the capacity (in the monochromatic case) as
function of the ratio $L/x_\mathsf{R}$, for $\bar{n}=4$. The figure refers
to a one-dimensional setting, in which information is encoded at the
object plane along a straight line of length $L$, and diffraction is
caused by an infinitely long slit of size $2R$. The solid line is
the exact value of the capacity computed by numerical evaluation of
the set of effective transmissivities. The dashed line indicates the
approximation for near-field limit ($L/x_\mathsf{R} \gg 1$). The
dotted line is the approximation for the far-field limit
($L/x_\mathsf{R} \ll 1$). The inset is a magnification of the far-field region.}
\label{regimes}
\end{figure}

To evaluate the capacity for a generic value of the ratio
$L/x_\mathsf{R}$, it is necessary to numerically diagonalize the
transfer matrix $T$. For the sake of simplicity, here we present an
example of lower dimensionality, in which information is
encoded at the object plane along an infinite strip of size $L$, and
diffraction is caused by an infinitely long slit of size $2R$. 
The analysis of such $1D$ system is analogous to the $2D$ one,
yielding the following expressions for the far-field and
near-field capacities. 
In the far-field limit we have
\begin{equation}\label{ff1D}
C_\mathrm{ff}^{(1D)} = g\left( \frac{2 L}{x_\mathsf{R}} \, \bar{n}
\right) \, ,
\end{equation}
and in the near-field limit
\begin{equation}\label{nf1D}
C_\mathrm{nf}^{(1D)} = \frac{2L}{x_\mathsf{R}} \, g \left( \frac{x_\mathsf{R}}{2L} \, \bar{n} \right) \, .
\end{equation}
Figure~\ref{transmissivities} shows the transmissivities for
several values of the ratio $L/x_\mathsf{R}$ for the
one-dimensional problem. 
Figure~\ref{regimes} shows the capacity as
function of the ratio $L/x_\mathsf{R}$, compared with the limiting
expressions of Eqs.~(\ref{ff1D}) and~(\ref{nf1D}).

\section{Enhanced quantum communication via optical refocusing}\label{Refocus}

For a fair comparison between the optical communication through the 
optical system and the free-space one, we consider the case of free-space 
communication under the hypotheses that light is emitted by an object of 
surface $L^2$, propagates by a distance $D=D_o+D_i=D_o(1+M)$, and is 
finally detected on a surface of size $(ML)^2$.
This setting is depicted in Fig.~\ref{optical}(b).
The light propagation, both in the classical and quantum regimes, is 
characterized by the Fresnel number associated to this setting, i.e., 
$\mathcal{F} = M^2 L^4/(\lambda D)^2$ (see \cite{Shapiro} and references therein). 
In the far-field limit, $\mathcal{F} \ll 1$, only one mode is transmitted, 
with a corresponding transmissivity equal to the Fresnel number. 
On the other hand, in the near-field limit, $\mathcal{F} \gg 1$,
$\mathcal{F}$ equals the number of modes which are transmitted
with unit transmissivity. For these two regimes we can hence compute
the free-space classical capacity according to~\cite{broadband}. 
By comparison with Eqs.~(\ref{farfield}), (\ref{nearfield}), it
follows that the presence of the radius-$R$ lens enhances the
classical capacity over the free-space propagation only for not too 
large $M$, i.e., for not too large imaging plane (specifically when 
$\pi R/x_\mathsf{R} > M/(1+M)$ in the far-field, and when 
$\pi^{1/2} R/L > M/(1+M)$ in the near-field limit). 
This rather counterintuitive effect originates from the simplifying 
assumption we made in writing Eq.~(\ref{transfer}) which implicitly 
states that the only light rays reaching the imagine plane are those 
which passes through the pupil, while the other are lost.  
Clearly the presence of such loss mechanism (which is not accounted for 
in the free-space calculation) is uninfluential as long as the image plane 
is small, while it becomes relevant for large imaging screen: 
this is exactly where the free-space starts to outperform the 
propagation through the lens.

Below we present a comparison between the performances 
of the refocusing apparatus, scenario (a), and those of the 
free-space propagation case, scenario (b).
The goal is to produce results which are not affected by the approximation 
we made in writing Eq.~(\ref{transfer}) (i.e., the fact that we have 
implicitly assumed that all photon which do not hint the lens 
will not be transferred on the imaging plane).
In order to do that, we also consider a third scenario, denoted by (c),
which is depicted in Fig.~\ref{optical}(c).

\subsection{Far-field regime for scenario (a)}\label{Refocus:ff}

Let us first consider the case in which the scenario (a) 
is operated in the far-field regime~(\ref{FARFIE}), which according to Eq.~(\ref{rayleigh}) corresponds to have 
\begin{eqnarray}\label{FFRA}
\lambda \gg \frac{LR}{D_o}\;.
\end{eqnarray}
As already seen in the previous section, under this condition  the scenario (a) is characterized 
by having a single mode transmitted from the object plane 
to the image plane with  
an effective transmissivity
\begin{eqnarray}\label{gghhjj}
\eta_{(a)} = \pi^2 (L/x_R)^4 = \pi^2 \left( \frac{LR}{\lambda D_o} \right)^4 \ll 1 \; .
\end{eqnarray}
In the scenario (b), the field propagates freely from the object 
plane to the image plane. 
In this case the far-field regime is equivalent to impose
\begin{eqnarray}\label{FFFP}
\lambda \gg \frac{LML}{D}= \frac{L^2}{D_o} \frac{M}{M+1} \; , 
\end{eqnarray}
which in principle is independent from the far-field 
condition~(\ref{FFRA}) for (a). 
Under condition~(\ref{FFFP}) also (b) will admit the propagation of a
single mode which is now attenuated by
\begin{eqnarray}\label{etab}
\eta_{(b)} = \pi \left( \frac{L_o L_i}{\lambda D} \right)^2 =
\pi \left(\frac{M}{M+1}\right)^2 \; \left( \frac{ L^2}{\lambda D_o} \right)^2 \ll 1 \; ,
\end{eqnarray}
where $L_o= L$ is the dimension of the object plane, $L_i=ML$ the dimension 
of the image plane, and $D=D_o+D_i$ is the distance between them.
Then, assuming that both~(\ref{FFRA}) and~(\ref{FFFP}) hold, the ratio between 
the transmissivities is
\begin{eqnarray}
r_1 =  \frac{\eta_{(a)}}{\eta_{(b)}}
= \pi \left(\frac{R^2}{\lambda D_o}\right)^2 \; \left( \frac{M+1}{M} \right)^2\;,
\end{eqnarray}
As already noticed this is not always larger than one: indeed there is a 
condition which $M, R, D_o, \lambda$ have to satisfy for this to happen. 
We now show that such condition is equivalent to impose that the loss 
induced by the pupil should be negligible. To do so consider the alternative 
scenario (c), in which the field propagate from the object plane to the 
image plane while instead of the lens we have just a hole in the absorbing 
screen of the pupil.
This configuration can be treated by analyzing separately 
the free-space propagation from the object plane to the absorbing screen 
$(o \rightarrow s)$ and the free-space propagation from the absorbing screen 
to the imaging plane $(s \rightarrow i)$. We notice that the condition Eq.~(\ref{FFRA}) 
imposes that both these propagations take place in the far-field regime~\cite{notaS1}.
Therefore we can conclude that, under the condition~(\ref{FFRA}), in the scenario (c) 
there will be a single propagating mode $(o\rightarrow s \rightarrow i)$.
It is attenuated by a factor
\begin{equation}
\eta_{(c)}^{(o\rightarrow s)} = \pi \left( \frac{ LR}{\lambda D_o} \right)^2
\end{equation}
in the section $(o\rightarrow s)$ and by 
\begin{equation}
\eta_{(c)}^{(s \rightarrow i)} = \pi \left( \frac{ RML}{\lambda D_i} \right)^2
= \pi  \left( \frac{ LR}{\lambda D_o} \right)^2
\end{equation}
in the section $(s \rightarrow i)$. 
The overall attenuation cannot be bigger than the product of the two factors, i.e.,
\begin{eqnarray}
\eta_{(c)} \leqslant \eta_{(c)}^{(o\rightarrow s)} \; \eta_{(c)}^{(s \rightarrow i)} =   \pi^2  \left( \frac{ LR}{\lambda D_o} \right)^4 = \eta_{(a)} \; .
\end{eqnarray}
Now, if the presence of the absorbing screen around the pupil 
has to be negligible, then we must have that the loss that it 
induces are equal to the one of free-space propagation. 
Since $\eta_{(c)} \leqslant \eta_{(a)}$, this implies that the 
regime in which we can neglect the effect of absorbing screen 
is exactly the one in which $r_1$ is greater than $1$. 
In other words the detrimental effects we see for $r_1 < 1$ 
simply correspond to the screen's absorption.
As a final remark we also observe that the condition $r_1 >1$ plus 
the far-field regime~(\ref{FFRA}) for (a) enforce the far-field 
regime~(\ref{FFFP}) for the free-space propagation, indeed,
\begin{eqnarray}
\frac{\lambda D_o}{L^2} \frac{M+1}{M} & =& \left( \frac{\lambda^2 D_o^2}{R^2L^2}\right) \left( \frac{R^2}{\lambda D_o} \right) \frac{M+1}{M} \nonumber  \\
& = & \left( \frac{\lambda^2 D_o^2}{R^2L^2}\right) \; \sqrt{\frac{r_1}{\pi}} > \left( \frac{\lambda^2 D_o^2}{R^2L^2}\right) \gg 1 \; . \label{FFFP1}
\end{eqnarray}

We now compare the performances of scenario (a) and (b) in terms 
of capacities by computing the ratio
\begin{eqnarray}
G_1 = \frac{C_{(a)}}{C_{(b)}} = \frac{g(\eta_{(a)} \bar{n})}{g(\eta_{(b)} \bar{n})} = \frac{g(r_1 \eta_{(b)} \bar{n})}{g(\eta_{(b)} \bar{n})} \; .
\end{eqnarray}
This is a monotonic function of $\bar{n}$, with $G_1 \simeq r_1$ for
$\bar{n} \ll 1$ (faint signal limit) and $G_1 \simeq 1$ for $\bar{n} \gg 1$
(semiclassical limit).
The enhancement in the transmission rate provided by
the optical refocusing system persists in the presence of
background thermal noise. In such a case, by encoding
classical information into coherent states, the expression
$g(\eta \bar{n})$ has to be replaced by 
$g(\eta \bar{n} + \bar{n}_\mathrm{th}) - g(\bar{n}_\mathrm{th})$,
where $\bar{n}_\mathrm{th}$ denotes the mean number of thermal background photons
per transmitted mode~\cite{HolWer}, yielding $G_1 \simeq r_1$ even in the 
very noisy limit $\bar{n}_\mathrm{th} \gg \bar{n}$.

\subsection{Near-Field regime for scenario (a)}\label{Refocus:nf}

The near-field regime for the scenario (a) is defined by the inequality~(\ref{NEARFIE}) which rewrites also as
\begin{eqnarray}\label{NFRA}
\lambda \ll \frac{LR}{D_o} \;.
\end{eqnarray}
From the discussion of Sec.~\ref{Mono} we know that in this regime the scenario (a) is characterized
by having a collection of $\nu_{(a)}$ modes which are perfectly 
transmitted from the object plane to the image plane,
\begin{eqnarray}
\nu_{(a)} = \pi (L/x_R)^2 = \pi \left( \frac{LR}{\lambda D_o} \right)^2 \gg 1\;.
\end{eqnarray}
In a similar way the near-field regime for the scenario (b) takes place when
\begin{eqnarray}\label{NFP}
\lambda \ll \frac{L_o L_i}{D}= \frac{L^2}{D_o} \frac{M}{M+1} \; , 
\end{eqnarray}
which is independent from the near-field condition for (a)~(\ref{NFRA}).
Under this condition also (b) admits $\nu_{(b)}$ modes which are 
perfectly transferred, with
\begin{eqnarray}
\nu_{(b)} = \pi \left( \frac{L_o L_i}{\lambda D} \right)^2 =
\pi \left(\frac{M}{M+1}\right)^2 \; \left( \frac{ L^2}{\lambda D_o} \right)^2 \gg 1 \; .
\end{eqnarray}
Assume then that both~(\ref{NFRA}) and~(\ref{NFP}) hold and define the quality factor
\begin{eqnarray}
r_2 := \frac{\nu_{(a)}}{\nu_{(b)}} = \left( \frac{M+1}{M} \right)^2 \; \left(\frac{R}{L}\right)^2\;.
\end{eqnarray}

As in the case of Sec.~\ref{Refocus:ff}, if there are no losses introduced
by the absorption of the rays propagating outside the lens, we
must have $r_2 \geqslant 1$.
To see this let us consider what happens in the scenario (c).
We first notice that the condition~(\ref{NFRA}) guarantees
that both the propagations $(o \rightarrow s)$ and $(s \rightarrow i)$ are in the
near-field regime. The number of modes that they allow for
perfect propagation is given by
\begin{eqnarray}
\nu_{(c)}^{(o \rightarrow s)} = \nu_{(c)}^{(s \rightarrow i)} = \pi \left( \frac{LR}{\lambda D_o} \right)^2 = \nu_{(a)} \;.
\end{eqnarray}
Notice that they are identical, due to the fact that $L_i/D_i= L/D_o$. 
This implies that $\nu_{(c)} \leqslant \nu_{(a)}$.
Then, it is clear that the presence of the pupil is negligible only when 
$\nu_{(c)}$ is larger than $\nu_{(b)}$, i.e., $r_2 \geqslant 1$, therefor proving the thesis.

We remark that the condition $r_2 \geqslant 1$, together with the 
near-field condition for the scenario (a) [Eq.~(\ref{NFRA})] is not sufficient 
to garantee the near-field condition for the scenario (b). 
Indeed we notice that
\begin{equation}
\frac{L^2}{\lambda D_o} \frac{M}{M+1} = \frac{LR}{\lambda D_o} \left( \frac{M}{M+1} \frac{L}{R} \right) \, ,
\end{equation}
which by itself does not imply the near-field condition for (b), i.e., 
Eq.~(\ref{NFP}).
That yields the possibility to have (a) in near-field and (b) in either 
near-field or far-field.

Let us examine the two cases.

\begin{itemize}

\item 
First assume that both (a) and (b) are in the near-field regime 
(of course under the constraint that $r_2 >1$ to exclude the 
absorption by the pupil). The ratio between the capacities becomes
\begin{eqnarray}
G_2 & = & \frac{C_{(a)}}{C_{(b)}} = \frac{\nu_{(a)} \; g(\bar{n}/\nu_{(a)})}{\nu_{(b)} \;g(\bar{n}/\nu_{(b)})} \nonumber \\
& = & r_2  \frac{g(\bar{n}/\nu_{(a)})}{g(r_2 \bar{n}/\nu_{(a)})} \, ,
\end{eqnarray}
which is a monotonic function of $\bar{n}$, with $G_2 \simeq 1$ for $\bar{n} \ll 1$
(faint signal limit) and $G_2 \simeq r_2$ for $\bar{n} \gg 1$ (semiclassical
limit).

\item 
Then assume that (a) is near-field and (b) is far-field, i.e.,
\begin{eqnarray}
\frac{L^2}{D_o} \frac{M}{M+1} \ll \lambda \ll \frac{LR}{D_o} \;.
\end{eqnarray}
Under these conditions one can immediately verify that the 
presence of the absorbing screen does not affect the performances 
of the scenario (b), and we can make a fair comparison.
In this case the gain becomes
\begin{equation}
G_3 = \frac{C_{(a)}}{C_{(b)}} = \frac{\nu_{(a)} \; g(\bar{n}/\nu_{(a)})}{ \;g(\eta_{(b)} \bar{n})} \; ,
\end{equation}
giving $G_3 \simeq 1/\eta_{(b)} \gg 1$ for $\bar{n} \ll 1$ (faint signal
limit) and $G_3 \simeq \nu_{(a)} \gg 1$ for $\bar{n} \gg 1$ (semiclassical limit).
In the presence of noisy thermal environment \cite{HolWer}, 
the expressions $g(\eta \bar{n})$ and $g(\bar{n}/\nu)$ have to be replaced by 
$g(\eta \bar{n} + \bar{n}_\mathrm{th}) - g(\bar{n}_\mathrm{th})$ and
$g(\bar{n}/\nu + \bar{n}_\mathrm{th}) - g(\bar{n}_\mathrm{th})$, respectively.
The advantages of optical refocusing hence persist even in the 
very noisy limit $\bar{n}_\mathrm{th} \gg \bar{n}$, yielding
$G_3 \simeq 1/\eta_{(b)}$.

\end{itemize}

\section{Communication with non monochromatic light}\label{NonMono}

Till now we have considered the case of monochromatic light,
we now want to compute the communication capacity of the optical 
system by assuming non-monochromatic light. 
An input signal over a time interval $T$ can be expanded in terms 
of monochromatic components at frequencies
$\omega_j = 2\pi j /T$, with $j=0,1,\dots \infty$. 
If the optical system is characterized by a certain bandwidth 
extending from $\Omega$ to $\Omega + \delta\Omega$, only a finite 
number of components are transmitted, corresponding to the frequencies
$\omega_n$ such that $\Omega \leqslant \omega_j \leqslant
\Omega+\delta\Omega$. Each monochromatic components contributes with
a term as in Eq.~(\ref{nearfield}), with $x_\mathsf{R}=\lambda_j
D_o/R=2\pi c D_o/(\omega_j R)$. In the following we will work under
the assumption that the frequency modes are either all in the
far-field regime or all in the near-field one.

\subsection{Far-field}

If all the transmitted frequency modes fulfill the far-field condition~(\ref{FARFIE}), we must have
\begin{eqnarray} \label{ffcond}
L \ll x_\mathsf{R} = 2\pi c D_o/(\omega_j R) \; ,
\end{eqnarray}
for all $\omega_j$. Consequently,
the results of Eq.~(\ref{farfield}) can be applied to the whole
spectrum. This allows us to derive the following expression for the
capacity
\begin{equation}
C_\mathrm{ff} = \sum_{\Omega \leqslant \omega_j \leqslant \Omega+\delta\Omega} g \left( \alpha \omega_j^4 \bar{n}_j \right) \, ,
\end{equation}
where
\begin{equation}
\alpha := \pi^2 \left( \frac{L R}{2\pi c D_0} \right)^4 \, ,
\end{equation}
and the parameter $\bar{n}_j$ counts the average number of photons at frequency $\omega_j$. 
If a mean power $P$ is employed, the parameters $\bar{n}_j$ ought to obey
the constraint
\begin{equation}
\frac{1}{T} \sum_{\Omega \leqslant \omega_j \leqslant \Omega+\delta\Omega} \; \hbar \omega_j \; \bar{n}_j = P \, .
\end{equation}
The maximization over photon number distributions satisfying the
input energy constraint can be worked out by Lagrange method,
yielding the optimal photon number distribution
\begin{equation}
\bar{n}_j = \left[ \alpha\omega_j^4\left(2^{\frac{\mu\hbar}{\alpha\omega_j^3 T}}-1 \right) \right]^{-1} \, ,
\end{equation}
with $\mu$ being the Lagrange multiplier. For sufficiently large
$T$ we can approximate the summations with integrals, and the
channel capacity reads
\begin{equation}
C_\mathrm{ff} = \frac{T}{2\pi} \int_\Omega^{\Omega+\delta\Omega} d\omega \, 
g\left( \frac{1}{2^{\frac{\mu\hbar}{\alpha T \omega^3}} - 1} \right) \, ,
\end{equation}
where the value of the Lagrange multiplier is determined by the
implicit equation
\begin{equation}
P = \frac{\hbar}{2\alpha\pi} \int_{\Omega}^{\Omega+\delta\Omega}
\frac{d\omega}{\omega^3} \;
\left(2^{\frac{\mu\hbar}{\alpha\omega^3 T}}-1 \right)^{-1} \, .
\end{equation}
A closed form for the classical capacity can be found in the
narrow-band limit, $\delta\Omega\ll\Omega$, in which we obtain the
approximate expression
\begin{equation}
C_\mathrm{ff}^\mathrm{nb} \simeq \frac{T\delta\Omega}{2\pi} g\left(\frac{2\pi P
\alpha \Omega^3}{\hbar \delta\Omega}\right) \, .
\end{equation}

\subsection{Near-field}

Let us now assume that all the frequency modes fulfill the
near-field condition~(\ref{NEARFIE}), i.e., 
\begin{eqnarray} \label{cod}
L \gg x_\mathsf{R} =2\pi c D_o/(\omega_j R) \; .
\end{eqnarray}
We can hence apply Eq.~(\ref{nearfield}) to the whole spectrum,
which allows us to write the channel classical capacity as follows:
\begin{align}
C_\mathrm{nf} & = \sum_{\Omega \leqslant \omega_j \leqslant \Omega+\delta\Omega}  \nu_j \, g \left( \frac{ \bar{n}_j }{\nu_j} \right) \\ 
& = \sum_{\Omega \leqslant \omega_j \leqslant \Omega+\delta\Omega} \beta \omega_j^2 \, g \left( \frac{ \bar{n}_j }{\beta \omega_j^2} \right) \, ,
\end{align}
where $\nu_j$ counts the number of transmitted modes of the
frequency $\omega_j$ [Eq.~(\ref{PU})], $\bar{n}_j$ counts the average number of photon of
that frequency, and where we have introduced
\begin{equation}\label{alpha}
\beta := \pi \left( \frac{L R}{ 2 \pi c D_o} \right)^2 \, .
\end{equation}
The optimization over the photon-number distribution under
the constraint of total power $P$ yields the optimal photon-number
distribution
\begin{equation}
\bar{n}_j = \frac{\beta \omega_j^2}{2^{\mu\hbar\omega_j/T}-1} \, ,
\end{equation}
with $\mu$ being the Lagrange multiplier. For sufficiently large
$T$ we can approximate the summations with integrals, and the
channel capacity reads
\begin{eqnarray}
C_\mathrm{nf} &=& \frac{\beta T}{2\pi} \int_{\Omega}^{\Omega+\delta\Omega}
d\omega \, \omega^2 \, g \left( \frac{1}{2^{\mu\hbar\omega/T}-1} \right) \nonumber \\
&=& \frac{\beta T}{2\pi q^3} \; \int_{q \Omega}^{q(\Omega+\delta\Omega)} d x \, x^2 \, 
g \left( \frac{1}{e^{x}-1} \right) \, ,
\end{eqnarray}
where we rescaled the Lagrange multiplier introducing the quantity
$q=\ln{(2)}\mu\hbar/T$, that is determined by the implicit equation
\begin{equation}\label{EConstraint}
P = \frac{\beta \hbar }{2\pi q^4 } \int_{q \Omega}^{q(\Omega+\delta\Omega)} d x \, \frac{ x^3}{e^{x}-1} \, .
\end{equation}
We can single out two limiting situations, the narrowband and the
broadband limits, for which analytical expressions can be obtained.

\subsubsection{Narrowband limit}

In the narrowband limit, $\delta\Omega \ll \Omega$, we obtain
\begin{equation}\label{nbc}
C_\mathrm{nf}^\mathrm{nb} \simeq  \frac{\beta}{2\pi} \; T \; 
\Omega^2 \delta\Omega \, g\left( \frac{2\pi {P}}{\beta \hbar \Omega^3 \delta\Omega} \right) \, .
\end{equation}
This can be casted in a more familiar form by noticing that in
this limit the power $P$ can be expressed as
\begin{equation}
P \simeq \frac{1}{T} \, \left( \frac{\delta\Omega T}{2\pi} \right) \, \hbar \Omega \, \bar{n}(\Omega) \, ,
\end{equation}
where ${\delta\Omega T}/{(2\pi)}$ is the total number of
frequencies and $\bar{n}(\Omega)$ is the density of mean photon 
number at frequency $\Omega$. Replacing this into
Eq.~(\ref{nbc}) and using $\beta \Omega^2=\pi(L/x_\mathrm{R})^2$,
where $x_\mathsf{R}$ is the Rayleigh length of the frequency
$\Omega$, we get
\begin{equation}
C_\mathrm{nf}^\mathrm{nb} \simeq \left( \frac{\delta\Omega T}{2\pi} \right) \, \pi \frac{L^2}{x_\mathrm{R}^2} \, g \left( \frac{x_\mathrm{R}^2}{\pi L^2} \, \bar{n}(\Omega) \right) \, .
\end{equation}
This expression shows that $C_\mathrm{nf}^\mathrm{nb}$ coincides with the single
frequency capacity~(\ref{nearfield}), multiplied by the total number
of frequencies $\delta\Omega T/(2\pi)$.

\subsubsection{Broadband limit}

In the broadband limit, we set $\Omega+\delta\Omega\to +\infty$, and we have 
\begin{eqnarray}
P & \simeq &  \frac{\beta \hbar }{2\pi q^4 } \; {\cal F}(q \Omega) \;, \\
C_\mathrm{nf}^\mathrm{bb} &\simeq& \frac{\beta T}{2\pi q^3} \;  {\cal G}(q \Omega)\;,
\end{eqnarray}
with ${\cal F} (z)$ and ${\cal G} (z)$ being the following
decreasing functions,
\begin{eqnarray} 
{\cal F} (z) &:=& \int_{z}^{\infty} d x \,
\frac{ x^3}{e^{x}-1} \, , \label{ff} \\
{\cal G}(z) &:=& \int_{z}^{\infty} d x \, x^2 \, g \left(
\frac{1}{e^{x}-1} \right) \, . \label{gg}
\end{eqnarray}
An analytical solution can be obtained by approximating $q\Omega$ to
zero. Notice that this is not formally correct if we want to
preserve the condition~(\ref{cod}) for all frequencies of the
spectrum: however since ${\cal F} (z)$ and ${\cal G} (z)$ are
smooth in the proximity of $z=0$ the error becomes
negligible small.
By close inspection of the Eqs.~(\ref{ff}) and (\ref{gg}) however,  
it follows that the approximations
$\mathcal{F}(z)\simeq\mathcal{F}(0)$,
$\mathcal{G}(z)\simeq\mathcal{G}(0)$ are justified if $z \lesssim
1$, i.e., for $q\Omega\lesssim 1$, which in turn implies:
\begin{equation}
\frac{2\pi}{\beta \hbar} \, \frac{P}{\Omega^4} \gtrsim
\mathcal{F}(0) \, .
\end{equation}
Under this assumptions it follows that the condition (\ref{cod}) is satisfied in the
semiclassical regime $P/(\hbar \Omega^2) \gg 1$.
Consequently 
we can write
\begin{eqnarray}
P &\simeq & \frac{\beta \pi^3 \hbar }{30 q^4 } \;, \\
C_\mathrm{nf}^\mathrm{bb} &\simeq& \frac{\beta \pi^3 T}{45 q^3} \;  \log_2 e\;,
\end{eqnarray}
which yields
\begin{equation}
C_\mathrm{nf}^\mathrm{bb} \simeq \frac{\beta \pi^2 T}{3\sqrt[4]{15}}
\left[ \frac{2 \pi P}{ \beta \hbar} \right]^{3/4}  \;  \log_2 e\;. \label{NEWLABEL}
\end{equation} 
These expressions are obtained by noticing that the
integrals in (\ref{ff}), (\ref{gg}) can be written in terms of the Bernoulli
numbers $B_k$ by means of the identity (see, e.g., \cite{intt})
\begin{equation}
\int_0^\infty dx \frac{x^{2n-1}}{e^{px}-1} =
(-1)^{n-1}\left(\frac{2\pi}{p}\right)^{2n} \frac{B_{2n}}{4n} \, .
\end{equation}
By comparing (\ref{NEWLABEL}) with the capacity of a multimode Gaussian bosonic
channel~\cite{yuen,broadband,QOchannel} we notice that the scaling
in $P$ is now changed (there it scales with as $P^{1/2}$). This is
due to the fact that in our case each frequency has multiple
degeneracies.

\section{Conclusion}\label{TheEnd}

We have computed the capacity of quantum optical communication
through an optical system characterized by a finite entrance
pupil. Our calculations provide general bounds on the efficiency
of quantum optical communication taking into account the effects
of light diffraction. This models a rather general situation in
long distance communication, where repeaters play the role of the
optical system used to refocus the signal. More generally, any
transfer of information which employ quantum degrees of freedom of light
--- from quantum key distribution~\cite{tools4,tools5,tools6} 
to quantum imaging~\cite{GLMS} as well as quantum discrimination 
problems~\cite{Qdiscr1,Qdiscr2,Qdiscr3,Qdiscr4} and quantum reading~\cite{Pirs,fu,QRC}
--- requires the propagation through an optical system, and it is
hence limited by diffraction.

We have argued that, when the optical system is used for information 
transmission, the effects of diffraction can be formally described as 
a quantum channel with correlated noise.
It follows that correlated noise may affect all the quantum information
protocols requiring propagation and detection of quantum light.

Finally, this formal equivalence has allowed us to apply tools that were developed in 
the framework of quantum memory channel characterization to the 
diffraction problem.
In particular, this has allowed us to show that, under certain conditions,
a converging optical apparatus can be used to achieve higher
transmission rates than the free-space field propagation. 
The tradeoff between loss and diffraction determines the conditions 
under which the intuitive benefits of optical refocusing can be rigorously proven.

\acknowledgments 
We acknowledge discussions with Lorenzo Maccone and Ciro Biancofiore. 
The research leading to these results has received
funding from the European Commission's seventh Framework Programme
(FP7/2007-2013) under grant agreements No.~213681, and by the
Italian Ministry of University and Research under the FIRB-IDEAS
project RBID08B3FM.  
V.G. also acknowledges the support of Institut Mittag-Leffler (Stockholm), 
where he was visiting while part of this work was done. 
S.P. acknowledges the support of EPSRC (EP/J00796X/1) and the European
Union (MOIF-CT-2006-039703).


\begin{thebibliography}{99}
\bibitem{notafs}
The expression in Eq.~(\ref{nearfield}) diverges in the limit of
$x_\mathsf{R} \to 0$. Clearly, that is an artifact due to the use of
the paraxial approximation ($R/D_o \ll 1$) in the derivation of the
capacity's formula. Indeed, the free-space propagation of light acts as a
filter which transmits only the transverse wave
vector components satisfying $|2\pi \mathbf{n}_o/L| < 2\pi/\lambda$ \cite{Goodman}. 
Thus the capacity of the optical system cannot be larger than 
$C = \frac{\pi L^2}{\lambda^2} \, g\left( \frac{\lambda^2 \bar{n}}{\pi L^2} \right)$.
\bibitem{notaS1}
Equation~(\ref{FFRA}) defines the far-field regime for the free-space 
$(o \rightarrow s)$ propagation. Similarly, since the ratio between the $(s \rightarrow i)$ 
distance $D_i$ and the dimension of the imaging plane $ML$ are equal to $D_o/L$ 
this condition defines also the far-field regime for the free-space $(s \rightarrow i)$ 
propagation.
\end{thebibliography}
\end{document}